\documentclass [12pt]{article}
\usepackage {graphicx}
\begin{document}

\title{The effect of noise on a hyperbolic strange
attractor in the system of two coupled van der Pol oscillators}

\author{Alexey Yu. Jalnine, Sergey P. Kuznetsov}

\maketitle

\begin{center}
\textit{Saratov Branch of Institute of Radio-Engineering and
Electronics, Russian Academy of Sciences, Zelenaya 38, Saratov,
410019, Russia} \vspace{2mm}
\end{center}

\begin{abstract}
We study the effect of noise for a physically realizable flow
system with a hyperbolic chaotic attractor of the Smale - Williams
type in the Poincar\'{e} cross-section [S.P. Kuznetsov, Phys. Rev.
Lett. 95, 2005, 144101]. It is shown numerically that slightly
varying the initial conditions on the attractor one can obtain a
uniform approximation of a noisy orbit by the trajectory of the
system without noise, that is called as the ``shadowing''
trajectory. We propose an algorithm for locating the shadowing
trajectories in the system under consideration. Using this
algorithm, we show that the mean distance between a noisy orbit
and the approximating one does not depend essentially on the
length of the time interval of observation, but only on the noise
intensity. This dependance is nearly linear in a wide interval of
the intensities of noise. It is found out that for weak noise the
Lyapunov exponents do not depend noticeably on the noise
intensity. However, in the case of a strong noise the largest
Lyapunov exponent decreases and even becomes negative indicating
the suppression of chaos by the external noise.
\end{abstract}

One of the intensively studied problems in nonlinear science is
the investigation of the effect of noise for the systems with
complex dynamical behavior, in particular for the ones possessing
strange chaotic attractors. It is known that strange attractors in
finite-dimensional nonlinear systems can be subdivided into three
main classes: uniformly hyperbolic, non-uniformly hyperbolic
(quasi-hyperbolic) and non-hyperbolic \cite{SSTC,AVOS,KH,DV,GH}.
Effects of noise on attractors of these classes have some specific
features \cite{AVOS}.

The objects referred to as non-hyperbolic attractors, or
quasiattractors, are not well defined. They are composed of a set
of complex orbits including chaotic limit sets and stable periodic
orbits with extremely narrow basins of attraction. In this case,
the presence of external noise appears as a saving remedy for
using the tools of the nonlinear dynamics based on existence of a
definite unique probabilistic measure. Thus, in principle, the
account of noise is of crucial significance for non-hyperbolic
attractors. For hyperbolic attractors the effect of noise is not
so essential because their intrinsic chaotic properties are
sufficient to ensure legitimacy of the description in terms of a
natural invariant measure. This statement is validated rigorously
on a solid axiomatic basis for the uniform hyperbolic attractors
(the so-called SRB measures by Sinai, Ruelle, and Bowen
\cite{SSTC,AVOS,KH,DV,GH,Young}). It is known that the SRB
measures correspond to zero-noise-limit. It means that they serve
as a good approximation for the systems under a weak noise as well
\cite{Young,Kifer}. Moreover, a strong result may be formulated in
concern to individual orbits basing on the so-called shadowing
lemma \cite{GH,FS,Bowen,LLWL,AKVSK}. Namely, on a time interval of
duration as long as wished, any motion of the system with weak
noise in the sustained regime may be represented approximately by
an orbit on the attractor of the system without noise. In this
sense, the noise may be regarded as unessential at all! As for the
non-uniform hyperbolic attractors, for conditions of existence of
invariant measures and for shadowing properties of orbits, we
refer a reader to a vast recent literature (see \cite{DV} and
references therein).

Formally, the understanding the effect of noise seems to be in the
clearest state for the uniformly hyperbolic attractors, but, in
fact, it is not yet a state that can satisfy a physicist. Indeed,
no physical examples of the uniformly hyperbolic attractors have
been known until recent time. For this reason, the formulation of
the problem in the physical aspect was not possible till now.

An idea of implementation of a kind of a uniformly hyperbolic
attractor was advanced in Ref. \cite{Kuz} in application to a
system of two coupled non-autonomous van der Pol oscillators. In
the Poincar\'{e} map of this system a chaotic attractor has been
found similar to the Smale - Williams solenoid. An analogous
system has been built as an electronic device and studied in
experiment \cite{KSel}. Numerical verification of conditions for a
theorem guaranteeing the hyperbolicity was performed in Ref.
\cite{KSat}. Some other examples of complex dynamics were
discussed in Refs. \cite{KPik,IJK,JK,KIO} basing on the same idea.

In the current article we report some results concerning the
effect of white Gaussian noise on the system of alternately
excited non-autonomous oscillators \cite{Kuz}. We discuss the
numerical simulation of the dynamics in the presence of noise and
present an algorithm for locating the shadowing trajectories,
which reproduce with a certain accuracy the orbits of the noisy
system. The algorithm is based on a step-by-step approach to the
noisy orbits and it gives a uniform approximation with a mean
deviation not depending upon the length of the time interval. We
demonstrate numerically that for the weak noise the magnitude of
the deviation from the noisy orbit remains small and it tends to
zero linearly with the decrease of the noise intensity.
Additionally, we analyze a dependence of the Lyapunov exponents on
the noise level. At weak noise they do not depend noticeably on
the noise intensity. This result agrees well with the assertion
that the SRB measure corresponds to the zero-noise limit of the
probabilistic measure. For strong noise, the largest Lyapunov
exponent decreases. It can become even negative indicating the
suppression of chaos by the external noise.

We study the model system determined by the following equations:
\begin{equation}
\begin{array}{ll}
\ddot{x} - [A \cos(\omega_{0}t/N) - x^{2}]\dot{x} + \omega_{0}^{2}x =
\varepsilon y \cos \omega_{0}t + D_{1}\xi(t), \\
\ddot{y} - [- A \cos(\omega_{0}t/N) - y^{2}]\dot{y} + 4
\omega_{0}^{2}y = \varepsilon x^{2} + D_{2}\xi(t).
\end{array}
\label{eq:e1}
\end{equation}
It is a pair of coupled non-autonomous van der Pol oscillators
with basic frequencies $\omega_0$ and $2\omega_0$. The control
parameters in both subsystems slowly vary periodically in time in
counter-phase $(\pm A \cos(\omega_0 t / N))$, and some special
type of coupling between the subsystems is introduced. The
frequency ratio $N$ is assumed to be an integer. A turn-by-turn
transfer of excitation between the subsystems is accompanied with
a transformation of the phase at successive periods of modulation
governed by the expanding circle map, or Bernoulli map:
$\varphi_{n+1}=2\varphi_n+const \pmod{2\pi}$. The gaussian white
noise $\xi(t)$ with $\langle\xi(t)\rangle=0$ and
$\langle\xi(t)\xi(t-\tau)\rangle=\delta(\tau)$ is added to the
right-hand parts of the equations. Parameters $D_{1,2}$
characterizing the noise intensity can be varied in a wide range.

As argued in the previous studies \cite{Kuz,KSel,KSat}, the
Poincar\'{e} map of the noiseless system defined for a period of
the external driving $T=2 \pi N / \omega_0$ possesses a uniformly
hyperbolic chaotic attractor, namely, a Smale - Williams solenoid
embedded in the four-dimensional phase space.

\begin{figure}[htbp]
\begin{center}
\includegraphics[width=6.5in]{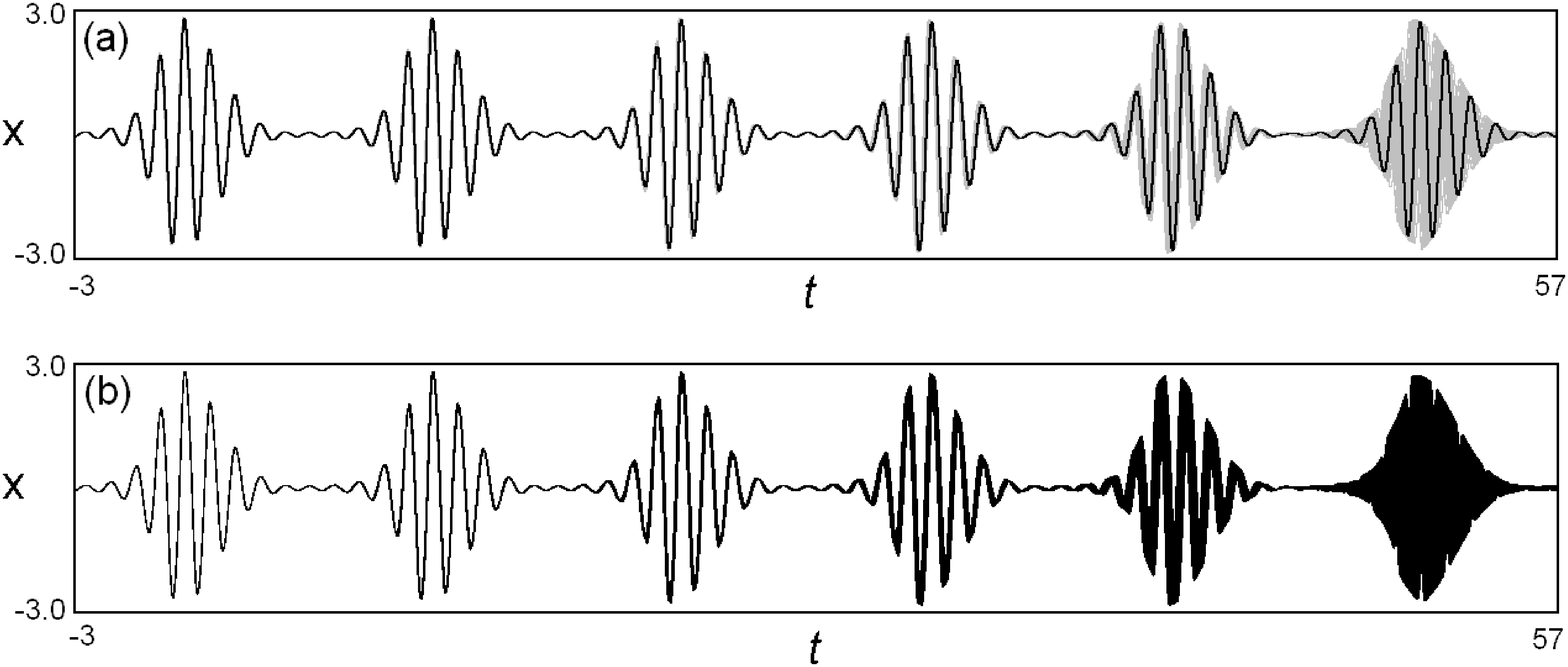}
\caption{Temporal realizations of the variable $x$ of the
model~(\ref{eq:e1}): (a)~superimposed 100 samples obtained at the
noise level $D_1=D_2=0.02$ (gray) and the trajectory without noise
(black), all with the same initial conditions; (b)~superimposed
100 samples for the system without noise ($D_1=D_2=0$) for the
ensemble of slightly different initial conditions. Here and
thereafter the parameters of the system~(\ref{eq:e1}) are chosen
to be $A=3.0, \varepsilon = 0.5, \omega_0 = 2\pi$, and $N = 10$.}
\label{fig:f1}
\end{center}
\end{figure}

For numerical solution of the stochastic equations~(\ref{eq:e1})
we exploit a second-order order method described in Ref.
\cite{MP}. A plot in Fig.~\ref{fig:f1}(a) shows the results of
computation for the noisy system at the parameters $A=3.0,
\varepsilon=0.5, \omega_0=2\pi, N=10$ and at the noise intensity
$D_1=D_2=0.02$. In gray color we show $100$ superimposed samples
of the process under effect of noise started from identical
initial conditions. Due to the noise, in the final part of the
interval of observation the states appear to be essentially
different because of instability intrinsic to the orbits on the
chaotic attractor. As the result, the picture becomes fuzzy. For
comparison, the curve shown in black color is related to the
system without noise, started from the same initial conditions.

It is easy to demonstrate qualitatively that a very similar
picture is observed in the noiseless system if one considers an
ensemble of samples with a slight deviation of the initial
conditions. In Fig.~\ref{fig:f1}(b) we show a set of $100$ samples
for the system without noise launched from initial conditions with
a small random variation near the same initial state as in the
previous diagram. The range of the random variations is specially
selected to obtain close degree of mutual divergence of the orbits
on the considered time interval in comparison with that produced
by the effect of noise.

\begin{figure}[htbp]
\begin{center}
\includegraphics[width=6.5in]{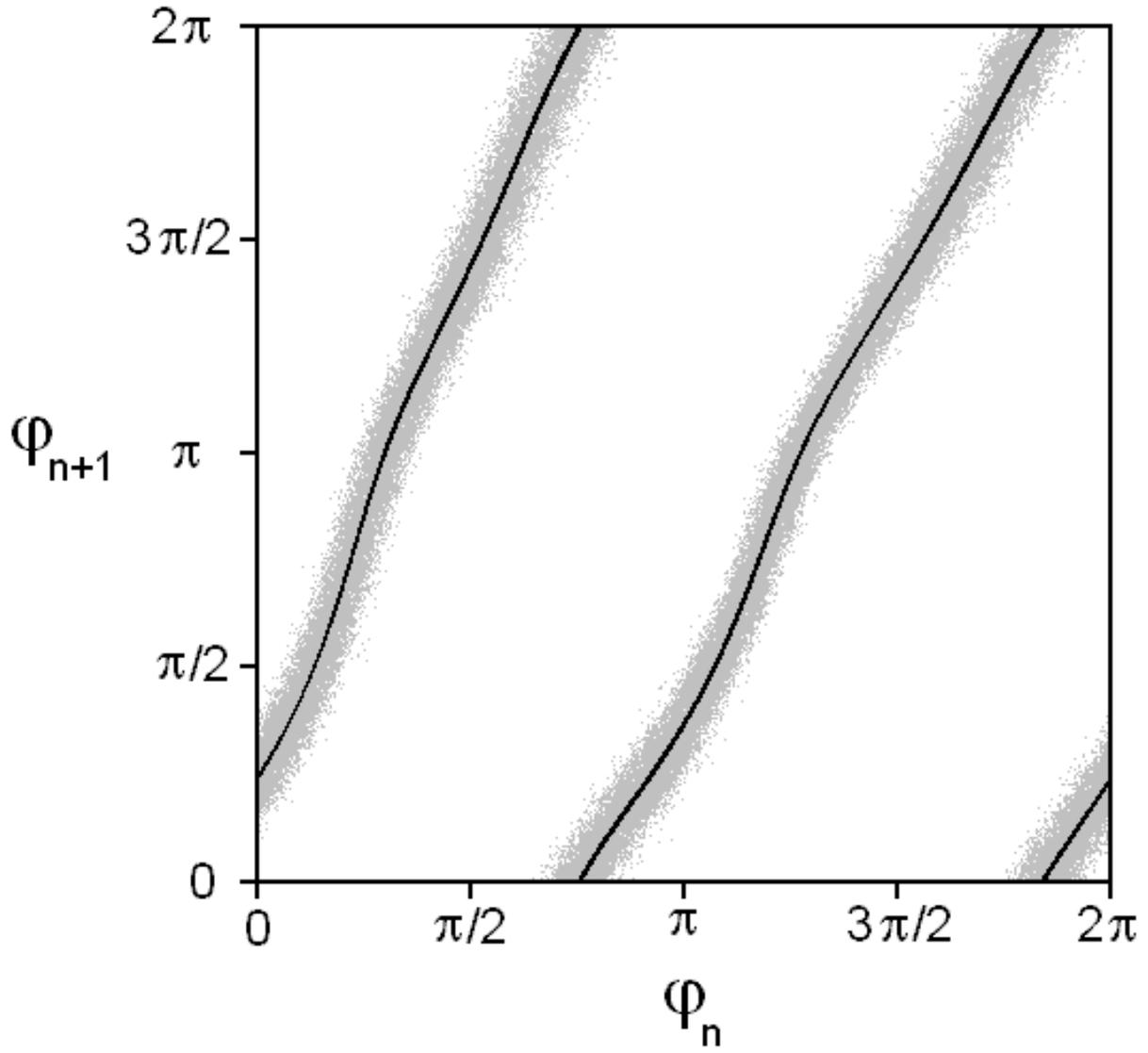}
\caption{ Iteration diagrams for the phase of the first subsystem
of the model~(\ref{eq:e1}) with noise $D_1=D_2=0.1$.}
\label{fig:f2}
\end{center}
\end{figure}

As shown for the noiseless system, the phases determined at
successive stages of activity for one of the subsystems obey
approximately the Bernoulli map. It is interesting to investigate
the influence of noise on the iteration diagrams for phases. Such
diagrams were used in Refs. \cite{Kuz,KSel,KSat,KPik} for
substantiation of the classification of the attractors as the
hyperbolic ones. Taking into account that during the active stage
the oscillations of $x(t)$ are close to sinusoidal ones with
modulated amplitude and floating phase ($x(t) \sim
\cos(\omega_0t+\varphi)$, see Ref. \cite{Kuz}), let us determine
the phase at the time moments $t_n=t_0+nT$ as follows:
\begin{displaymath}
\varphi_n=\arg[\dot{x}(t_n)+i\omega_0 x(t_n)].
\end{displaymath}
Figure~\ref{fig:f2} shows the iteration diagrams for the phases in
presence of noise (gray) and without noise (black). Note, that the
presence of noise of sufficiently low intensity does not change
the topological nature of the phase map, which remains in the same
class as the Bernoulli map $\varphi_{n+1}=2\varphi_n+const
\pmod{2\pi}$.

\begin{figure}[htbp]
\begin{center}
\includegraphics[width=6.5in]{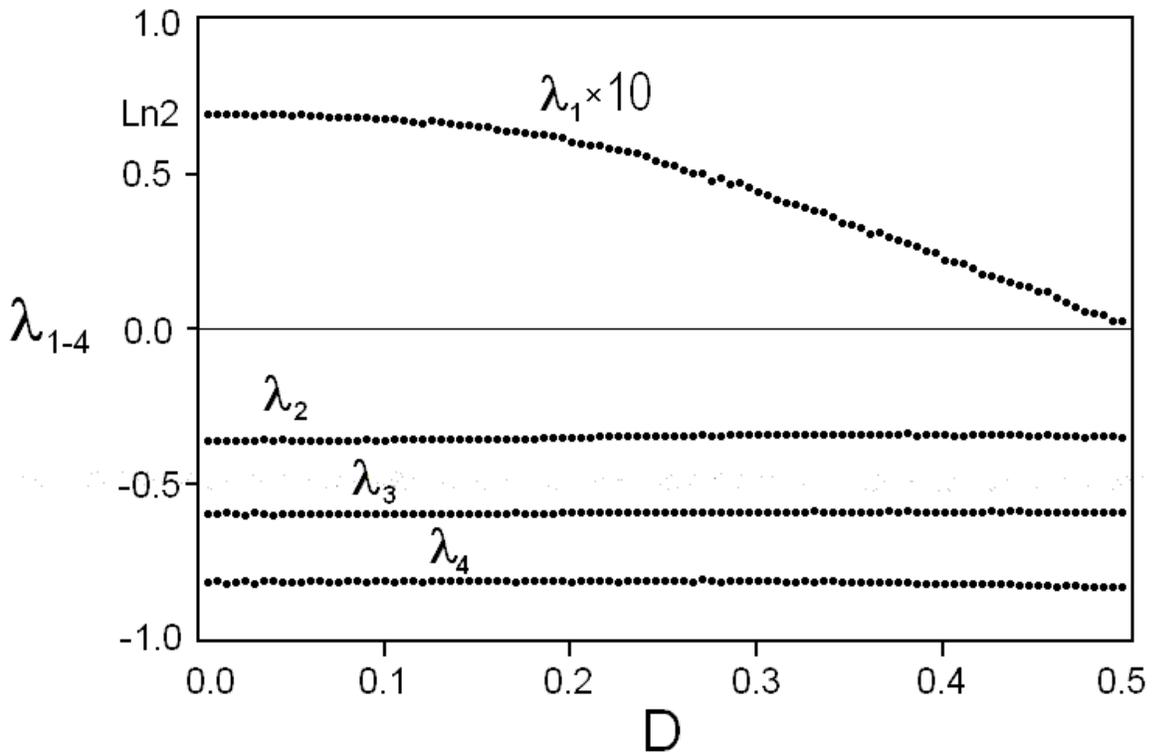}
\caption{Spectrum of the Lyapunov exponents versus intensity of
noise for the model~(\ref{eq:e1}) with $D_1=D_2=D$.}
\label{fig:f3}
\end{center}
\end{figure}

Another approach that allows us to compare the noisy and the
noiseless dynamics is based on the analysis of the Lyapunov
exponents. For our model they can be computed from the linearized
equations
\begin{equation}
\begin{array}{ll}
\ddot{\tilde{x}} - [A \cos(\omega_{0}t/N) - x^{2}]\dot{\tilde{x}}
+ (2x\dot{x} + \omega_{0}^{2})\tilde{x} =
\varepsilon \tilde{y} \cos \omega_{0}t, \\
\ddot{\tilde{y}} + [A \cos(\omega_{0}t/N) + y^{2}]\dot{\tilde{y}}
+ (2y\dot{y}+4\omega_{0}^{2})\tilde{y} = 2 \varepsilon x
\tilde{x},
\end{array}
\label{eq:e2}
\end{equation}
where tilde designates small perturbations of the dynamical
variables. These equations have to be solved numerically together
with the stochastic equations~(\ref{eq:e1}). To obtain the
spectrum of all the four Lyapunov exponents we consider a set of
four perturbation vectors
$(\tilde{x},\dot{\tilde{x}}/\omega_0,\tilde{y},\dot{\tilde{y}}/2\omega_0)$
and apply the procedure of Gram - Schmidt orthogonalization after
each period of the parameter modulation. In Fig.~\ref{fig:f3} the
Lyapunov exponents are plotted versus the noise intensity
parameter $D=D_1=D_2$. At small intensities of noise, the largest
Lyapunov exponent is close to the value of $T^{-1}\ln 2$, which
corresponds to the approximate description of the dynamics by the
Bernoulli map. A notable deflection appears only at a sufficiently
high level of noise, namely, at $D \geq 0.2$. At $D \sim 0.5$ the
effect of noise is already very relevant, the largest Lyapunov
exponent crosses zero and becomes negative indicating the
suppression of the intrinsic dynamical chaos by the external
noise. Dependence of other Lyapunov exponents on the noise
intensity is not noticeable at all.

Now we turn to the main part of the present paper. Namely, we are
going to illustrate numerically that in our model with a
hyperbolic attractor the weak noise is indeed non-essential, i.e.,
a typical noisy orbit can be reproduced for a long time interval
by a trajectory without noise due to a careful appropriate choice
of the initial conditions. This statement follows mathematically
from the hyperbolic nature of the attractor and is based on the
applicability of the shadowing lemma.

Suppose two trajectories are launched form identical initial
conditions, one in the ``pure'' system without noise and another
one in the system with noise. Obviously, the ``noisy'' orbit will
diverge from the ``pure'' one. Now, let us try to vary the initial
conditions for the pure trajectory to get an approximation for the
noisy orbit in the best way at a long time interval. The whole
construction is performed for a definite sample of the noisy orbit
obtained with the same sample of noise.

\begin{figure}[htbp]
\begin{center}
\includegraphics[width=6.5in]{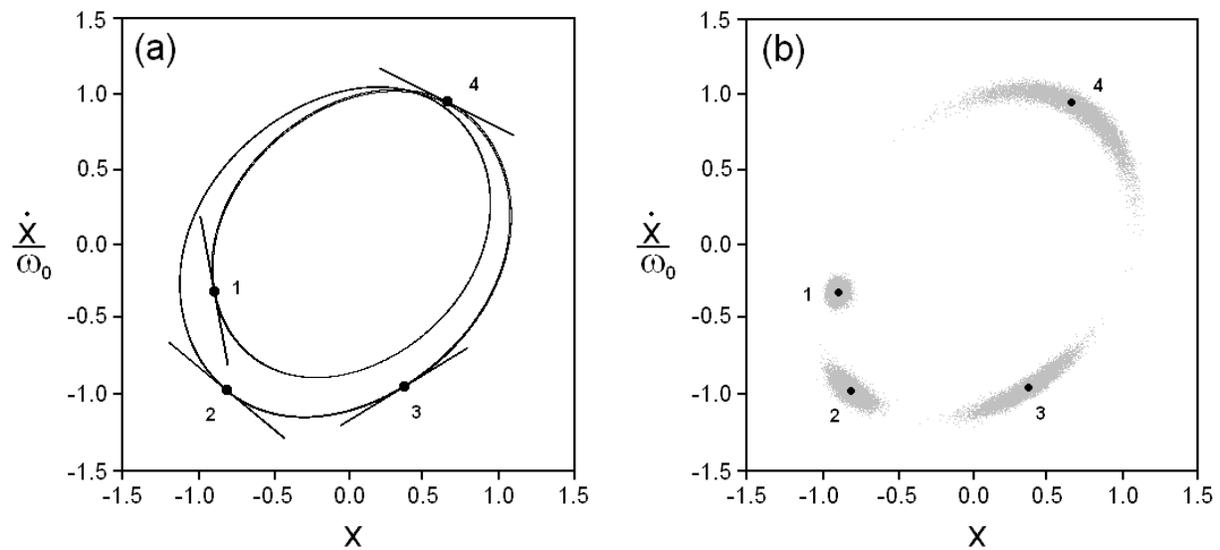}
\caption{(a)~The stroboscopic cross-section of the attractor of
the system~(\ref{eq:e1}) at $t_0=0.0$. The attractor is projected
on the phase plane of the first subsystem. (b)~Evolution of the
cloud of representative points launched from identical initial
conditions in the presence of noise ($D_1=D_2=0.02$). The figures
$1 \ldots 4$ indicate the number of steps of the Poincar\'{e} map.
An ensemble of $10^4$ orbits is shown. Black dots correspond to
the system without noise starting from the same initial
conditions.} \label{fig:f4}
\end{center}
\end{figure}

To explain the method of selection of the initial conditions, it
seems appropriate to consider the Poincar\'{e} map produced by a
period-$T$ stroboscopic section of the flow system~(\ref{eq:e1})
without noise. Let us suppose that we have an instantaneous state
given by a vector ${\bf
V}_n=(x,\dot{x}/\omega_0,y,\dot{y}/2\omega_0)$ at $t_n=t_0+nT$.
Then, after the time interval $T$ we have a new state
\begin{equation}
{\bf V}_{n+1}=\hat{{\bf F}}_{t_0}({\bf V}_n).
 \label{eq:e3}
\end{equation}
In practice, such map can be obtained from numerical integration
of the system~(\ref{eq:e1}) with $D_1=D_2=0$. In
Fig.~\ref{fig:f4}(a) one can see a portrait of the attractor of
the map projected onto the plane $(x,\dot{x}/\omega_0)$. The
attractor manifests filaments of an infinite number of wraps
possessing the Cantor-like structure in the cross-section. Solid
black dots in the picture denote four successive points of the
stroboscopic section of one specially chosen trajectory on the
attractor. The black line passing through each dot designates the
respective unstable direction ${\bf D}^u$, which is tangent to the
unstable manifold $W^u$ at the given point. These directions can
be simply approximated numerically from long-time evolution of an
arbitrarily chosen unit perturbation vector $\tilde{{\bf
V}}_n=(\tilde{x},\dot{\tilde{x}}/\omega_0,\tilde{y},\dot{\tilde{y}}/2\omega_0)$
at $t_n=t_0+nT$, since such a vector governed by the linearized
system~(\ref{eq:e2}) tends to the unstable direction associated
with a single positive Lyapunov exponent.

Now one should take a note of how a cloud of representative points
evolves from the same initial conditions in the presence of noise.
An illustration is shown in Fig.~\ref{fig:f4}(b). The dots
pictured in gray are obtained in the Poincar\'{e} cross-section,
i.e., stroboscopically at each next period $T$ from $10^4$ sample
orbits of the noisy system at $D_1=D_2=0.02$. The black dots
correspond to the trajectory of the system without noise
($D_1=D_2=0$) launched from the same initial conditions. Note,
that the cloud stretches along the unstable direction tangent to
the filaments forming the attractor, but it does not grow in the
radial direction.

The algorithm for localization of the ``pure'' trajectory, which
approximates a noisy orbit, consists in the following. Let us
denote the original noisy orbit as ${\bf V}_{noisy}(t)$, while the
pure trajectory in zero approximation is denoted as ${\bf
V}^{(0)}(t)$. We start first from the initial conditions at
$t=t_0$ identical for the noisy and the pure orbit: ${\bf
V}_{noisy}(t_0)={\bf V}^{(0)}(t_0)$. The starting point is
supposed to belong to the attractor of the system without noise.
Then, computing the two trajectories ${\bf V}_{noisy}(t)$ and
${\bf V}^{(0)}(t)$, we consider the norm of the difference vector
$\parallel \Delta {\bf V}^{(0)}(t_1) \parallel =
\parallel {\bf V}_{noisy}(t_1)-{\bf V}^{(0)}(t_1) \parallel$ at
the time moment $t_1=t_0+T$. Next, we slightly modify the initial
condition for the pure orbit from ${\bf V}^{(0)}(t_0)$ to ${\bf
V}^{(1)}(t_0)$ with the purpose to minimize the norm of the
difference $\parallel{\bf V}_{noisy}(t_1)-{\bf
V}^{(1)}(t_1)\parallel$. It is done by variation of the phase of
the partial oscillator active at the time moment $t=t_0$. For our
system it corresponds  to variation of the initial state along the
unstable direction associated with the point on the attractor, or,
that is the same, along a filament of the attractor containing the
initial point.

In detail the procedure of the search for new initial conditions
${\bf V}^{(1)}(t_0)$ looks as follows. We define a set of $2m+1$
initial conditions stretched along the unstable direction ${\bf
V}_{1,k}(t_0)={\bf V}^{(0)}(t_0)+\Delta V^{(1)}(k/m){\bf D}^{u}$,
where $k=-m, \ldots, m$, $\parallel {\bf D}^{u}\parallel$=1, and
the maximum variation is chosen from the relation $\Delta
V^{(1)}=\parallel \Delta {\bf V}^{(0)}(t_1)
\parallel \exp {(-\lambda_{1}T)}$. We trace then $2m+1$ trajectories
till $t=t_1$ and choose the one at some $k=k_1$, which minimizes
the error $\parallel{\bf V}_{noisy}(t_1)-{\bf
V}_{1,k_1}(t_1)\parallel$. This trajectory is denoted as ${\bf
V}^{(1)}_1={\bf V}_{1,k_1}$. Next, we redefine the set of $2m+1$
initial conditions as ${\bf V}_{2,k}(t_0)={\bf
V}^{(1)}_{1}(t_0)+\Delta V^{(1)}(k/m^2){\bf D}^{u}$ and select the
trajectory, which minimizes the error $\parallel{\bf
V}_{noisy}(t_1)-{\bf V}_{2,k_2}(t_1)\parallel$ at $k=k_2$. It is
the trajectory ${\bf V}^{(1)}_2={\bf V}_{k_2}$. We repeat this
procedure of successive adjustment of the initial conditions again
and again obtaining a sequence of the initial conditions $\{{\bf
V}^{(1)}_{l}(t_0)\}_{l=1,2, \ldots}$ and estimate the limit ${\bf
V}^{(1)}(t_0)$. In computations, the procedure is stopped after a
sufficiently large number of steps, when further increase of
accuracy does not result in a decrease of the final deviation of
the pure trajectory from the noisy one at $t=t_1$.

Suppose we take a new initial condition for the pure orbit ${\bf
V}^{(1)}(t_0)$. Now with the same sample of the noise we again
trace two trajectories, the noisy one starting from ${\bf
V}_{noise}(t_0)$ and the pure one starting from ${\bf
V}^{(1)}(t_0)$ for a longer time interval up to $t_2=t_0+2T$, and
obtain the norm of the difference vector $\parallel \Delta {\bf
V}^{(1)}(t_2)\parallel = \parallel {\bf V}_{noisy}(t_2)-{\bf
V}^{(1)}(t_2)\parallel$. Then, modifying again the initial
condition for the pure orbit by variation of the initial phase of
the active oscillator, at this time in closer neighborhood of
${\bf V}^{(1)}(t_0)$, we try to minimize $\parallel{\bf
V}_{noisy}(t_2)-{\bf V}^{(2)}(t_2)\parallel$ and obtain the new
initial condition ${\bf V}^{(2)}(t_0)$. The procedure of variation
of the initial conditions is the same as the described above with
the only difference that the maximum variation $\Delta V^{(2)}$ is
chosen from the relation of $\Delta V^{(2)}=\parallel \Delta {\bf
V}^{(1)}(t_2)\parallel \exp {(-2 \lambda_{1}T)}$. Step by step we
successively increase the time interval $nT$ and select the
initial conditions for the pure orbit ${\bf V}^{(n)}(t_0)$, which
deliver minimal values for the norms of the differences
$\parallel{\bf V}_{noisy}(t_n)-{\bf V}^{(n)}(t_n)\parallel$. Note,
that the maximum variation of the initial condition at the $n$-th
step of the algorithm decays as $\Delta V^{(n)} \sim \exp{(-n
\lambda_1 T)}$. Due to the recurrent nature of the algorithm, it
appears that the approximation of the noisy orbit by the pure one
holds uniformly along the whole time interval $[t_0,t_0+nT]$.

\begin{figure}[htbp]
\begin{center}
\includegraphics[width=6.5in]{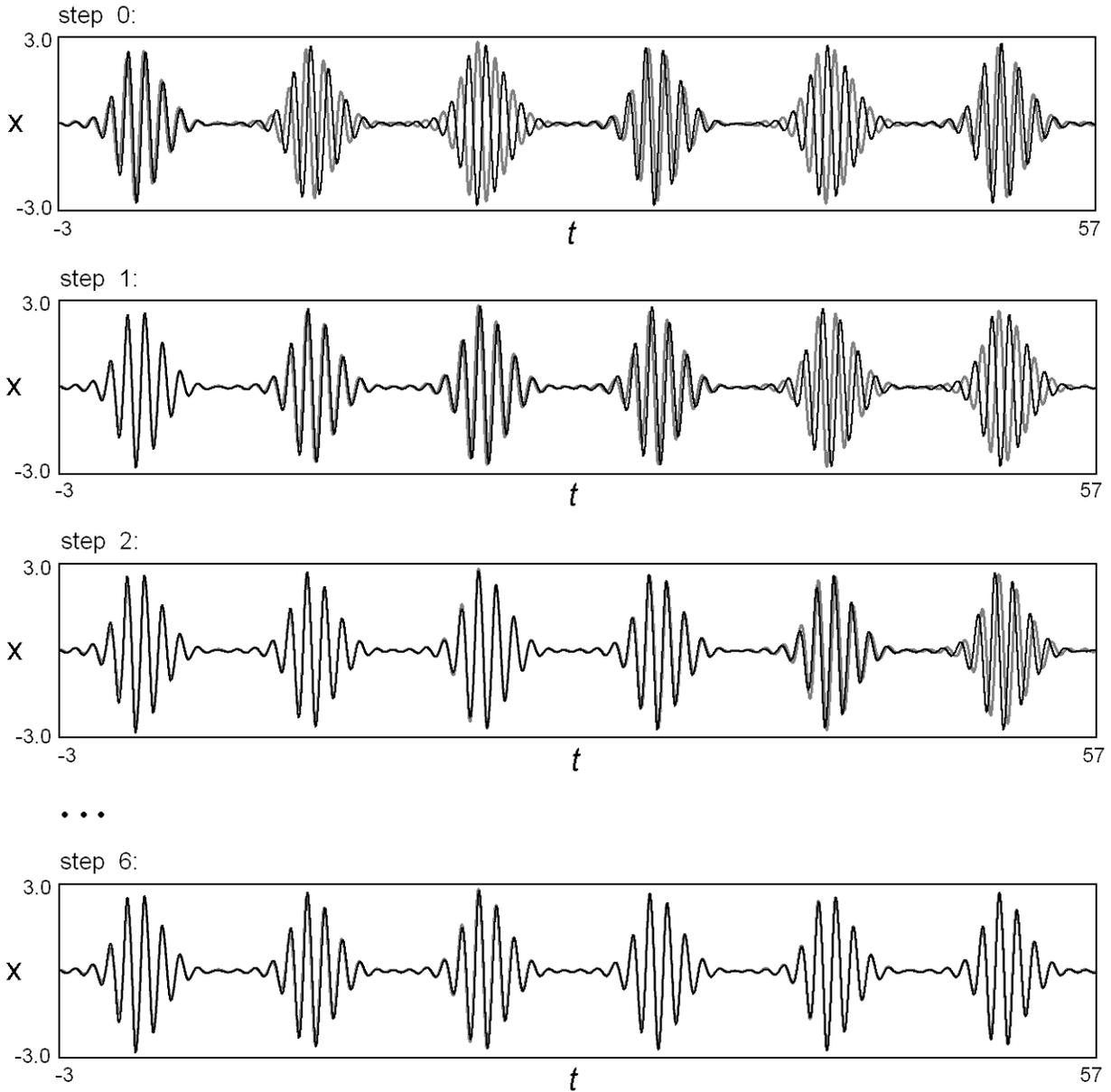}
\caption{Successive steps of constructing the shadowing trajectory
(black) to the given noisy orbits (gray). At the initial step both
orbits start from the identical initial conditions. At the last
presented diagram the shadowing takes place over the time range of
six modulation periods.} \label{fig:f5}
\end{center}
\end{figure}

Figure~\ref{fig:f5} illustrates results of application of several
steps of the above algorithm. Parameters of the system are taken
the same as those in the previous examples of computations, and
the noise intensity is $D_1=D_2=0.1$.

\begin{figure}[htbp]
\begin{center}
\includegraphics[width=6.5in]{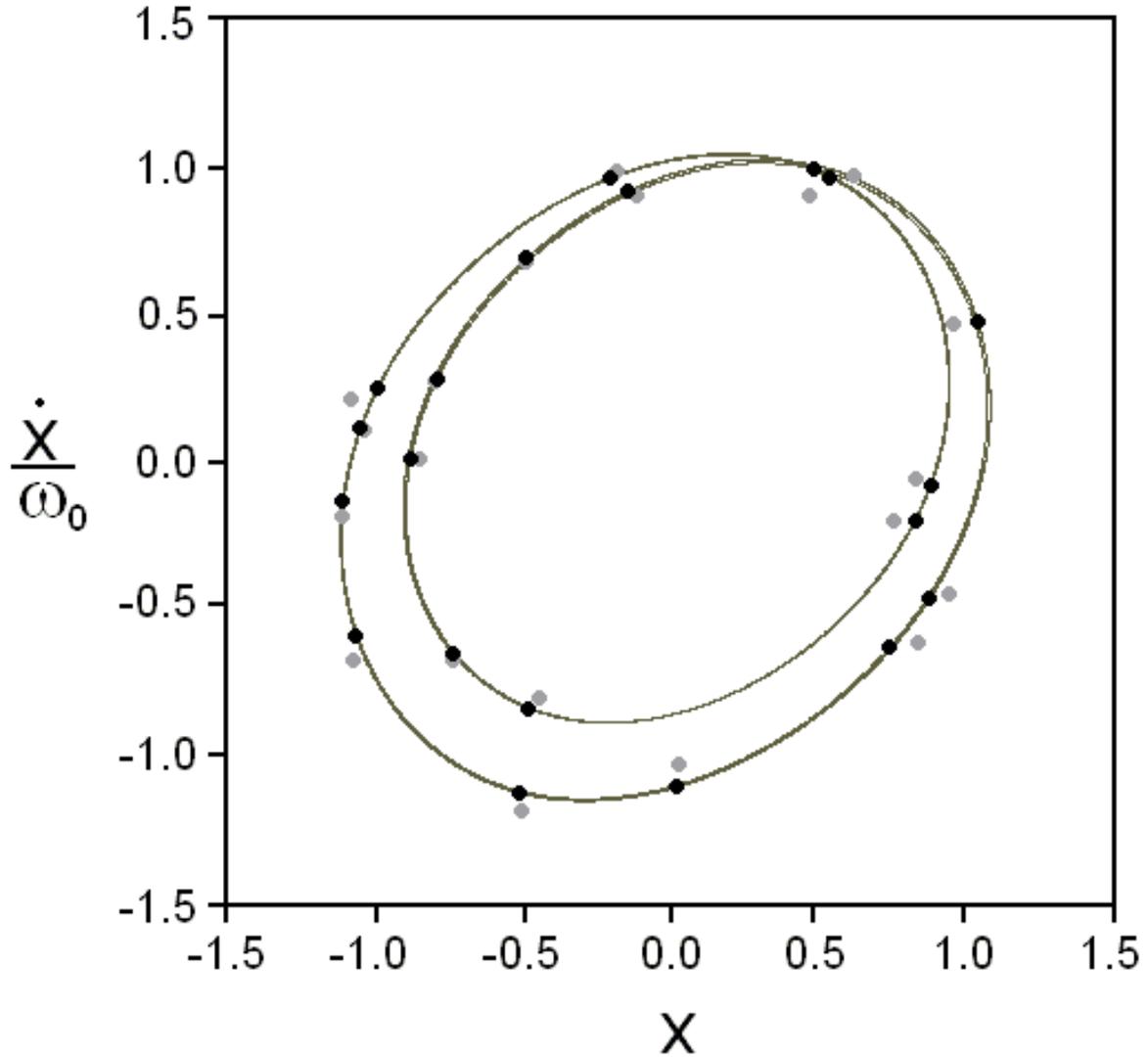}
\caption{Stroboscopic cross-section of the noisy (light gray) and
the shadowing noiseless (black) trajectories on a background of
the attractor (dark gray). The trajectories are projected onto the
phase plane of the first subsystem. The noise intensity
$D_1=D_2=0.04$.} \label{fig:f6}
\end{center}
\end{figure}

The first plot in Fig.~\ref{fig:f5} corresponds to the initial
step of the algorithm: the noisy (gray) and pure (black)
trajectories start from identical initial conditions. One can see
their sufficiently fast divergence: the phase synchronism
disappears already after one period of the parameter modulation
$T$. After the first modification of the initial conditions for
the pure orbit the divergence is delayed and the phase synchronism
persists over $1-2$ periods of $T$, but then the orbits diverge.
At the next steps of the algorithm the time intervals of existence
of the phase synchronism become longer and longer and occupy
finally the whole range in the diagram. Figure~\ref{fig:f6} shows
in gray color 20 dots of the projection of the noisy orbit at
$D_1=D_2=0.04$ on the $(x,\dot{x}/\omega_0)$ plane. In black color
$20$ dots are shown in the same stroboscopic cross-section of the
approximating pure trajectory recovered by the above method.

\begin{figure}[htbp]
\begin{center}
\includegraphics[width=6.5in]{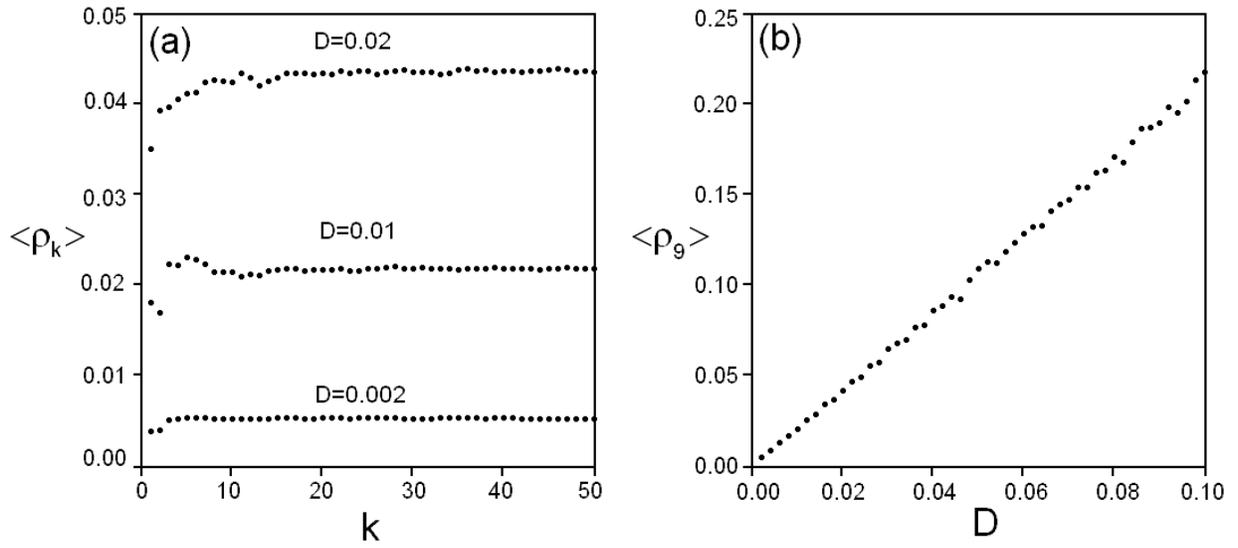}
\caption{(a)~Mean average mutual deviation of noise and shadowing
trajectories $\langle \rho_k \rangle$  versus the number of
modulation periods taken for the computations at the three noise
levels $D_1=D_2=0.002$, $0.01$ and $0.02$. (b)~The dependence of
$\langle \rho_k \rangle$ on the noise intensity $D$ for $k=9$.}
\label{fig:f7}
\end{center}
\end{figure}

To characterize the degree of closeness of a noisy orbit to a
shadowing pure orbit, we use the following value
\begin{displaymath}
\rho_k = \frac{1}{kT} \int\limits_{t_0}^{t_0+kT} \parallel{\bf
V}_{noisy}(t)-{\bf V}^{(k)}(t)\parallel dt,
\end{displaymath}
where $k$ designates the duration of the considered time interval
in units of the modulation period $T$. Averaging this quantity
over an ensemble of initial conditions and noise samples we obtain
a mean deviation $\langle \rho_k \rangle$, characterizing the
degree of closeness of the noisy and shadowing trajectories on the
attractor. In Fig.~\ref{fig:f7}(a) we present plots of the mean
deviation $\langle \rho_k \rangle$ versus the value of $k$ for
different noise levels. It can be seen in the figure that the mean
deviation depends on $k$ very weakly in the range $k = 8 \ldots
50$. At $k \sim 50-60$ the errors become noticeable due to the
finite-digital arithmetic and for the considered level of accuracy
further observation of the shadowing becomes impossible.

Figure~\ref{fig:f7}(b) shows a plot of the mean deviation $\langle
\rho_k \rangle$ computed for $k = 9$ in dependence on the
parameter of the noise intensity $D=D_1=D_2$. For each $D$ value
we consider a set of $100$ orbits launched from different initial
conditions and calculated for different noise series. Each noisy
orbit was approximated by the shadowing one with the method
described above, and the mean value $\langle \rho_9 \rangle$ was
obtained over the set of orbits. As seen from the figure, the
dependence $\langle \rho_9 \rangle$ versus $D$ looks like a linear
one in the range of $D$ from zero to $0.1$.

In this article we have examined the effect of Gaussian noise on a
physically realizable system with a uniformly hyperbolic
attractor. In the context of the shadowing lemma known from the
mathematical literature, one could expect that in our model the
effect of weak noise can be compensated by a careful selection of
initial conditions. We developed the numerical algorithm that
allows one to locate the shadowing trajectories of the noiseless
system and provide a uniform approximation of orbits of the noisy
system. Using this algorithm, we have demonstrated numerically
that the mean mutual deviation between the noisy orbit and the
noiseless shadowing trajectory does not depend noticeably on the
length of the considered time interval, but depends on the
intensity of noise. This dependence was shown to be almost linear
in a sufficiently wide range of the parameter $D$ variation. As
well, we have demonstrated that the weak noise does not change the
nature of the chaotic phase dynamics. Similarly, the weak noise
does not change noticeably the Lyapunov exponents characterizing
quantitatively the degree of instability of the motion. At larger
intensities the noise can suppress the intrinsic chaotic dynamics
of the system, and the largest Lyapunov exponent becomes negative.

The authors thank Dr. D.S. Goldobin for useful discussions and Dr.
M.D. Prokhorov for assistance in preparation of the paper. We
acknowledge support from the RFBR - DFG grant No. 04-02-04011.

\newpage

\end{document}